\documentclass[twocolumn,aps,superscriptaddress,showpacs]{revtex4}
\usepackage{amssymb}
\usepackage{amsmath}
\usepackage{graphicx}
\usepackage[normalem]{ulem}
\usepackage[dvips]{color}

\renewcommand\sout{\bgroup \color{red} \ULdepth=-.5ex \ULset}

\begin{document}

\title{Effect of the momentum dependence of nuclear symmetry potential on $\pi^{-}/\pi ^{+}$ ratio in heavy-ion
collisions}
\author{Yuan Gao}
\affiliation{School of Information Engineering, Hangzhou Dianzi
University, Hangzhou 310018, China}
\affiliation{School of Nuclear Science and Technology, Lanzhou
University, Lanzhou 730000, China}
\author{Lei Zhang}
\affiliation{School of Information Engineering, Hangzhou Dianzi
University, Hangzhou 310018, China}
\author{Hong-Fei Zhang}
\author{ Xi-Meng Chen}
\affiliation{School of Nuclear Science and Technology, Lanzhou
University, Lanzhou 730000, China}
\author{Gao-Chan Yong}\affiliation{Institute of Modern Physics, Chinese Academy of
Sciences, Lanzhou 730000, China}

\begin{abstract}
In the framework of the isospin-dependent
Boltzmann-Uehling-Uhlenbeck transport model, effect of the
momentum dependence of nuclear symmetry potential on $\pi^{-}/\pi
^{+}$ ratio in the neutron-rich reaction $^{132}$Sn+$^{124}$Sn at
a beam energy of $400$ MeV/nucleon is studied. We find that the
momentum dependence of nuclear symmetry potential affects the
compressed density of colliding nuclei, numbers of produced
$\pi^{-}$ and $\pi ^{+}$, as well as the value of $\pi^{-}/\pi
^{+}$ ratio. The momentum dependent nuclear symmetry potential
increases the compressed density of colliding nuclei, numbers of
produced resonances $\Delta(1232)$, $N^{*}(1440)$, $\pi^{-}$ and
$\pi ^{+}$, as well as the value of $\pi^{-}/\pi ^{+}$ ratio.
\end{abstract}

 \maketitle


Nowadays pion production in heavy-ion collisions has attracted
much attention in the nuclear physics community
\cite{xiao09,ditoro1,xu09,Rei07,yong06,yong101,yong102}. The
reason for this is that pion production is connected with the
high-density behavior of nuclear symmetry energy \cite{LiBA02}.
The latter is crucial for understanding many interesting issues in
both nuclear physics and astrophysics
\cite{Bro00,Dan02a,Bar05,LCK08,Sum94,Lat04,Ste05a}. The high-%
density behavior of nuclear symmetry energy, however, has been
regarded as the most uncertain property of dense neutron-rich
nuclear matter \cite{Kut94,Kub99}. Many microscopic and/or
phenomenological many-body theories using various interactions
\cite{Che07,LiZH06} predict that the symmetry energy increases
continuously at all densities. However, other models
\cite{Pan72,Fri81,Wir88a,Kra06,Szm06,Bro00,Cha97,Sto03,Che05b,Dec80,MS,Kho96,Bas07,Ban00}
predict that the symmetry energy first increases to a maximum and
then may start decreasing at certain suprasaturation densities.
Thus, currently the theoretical predictions on the symmetry energy
at suprasaturation densities are extremely diverse. Therefore,
what is crucially needed for the first step is a qualitative
observable to probe whether the symmetry energy at high densities
is soft or stiff, this has been done originally by qualitatively
comparing the n/p ratio of light and heavy preequilibrium emitting
clusters \cite{yong105}. Making further progress in determining
the symmetry energy at suprasaturation densities needs some
guidance from dialogues between experiments and transport models,
which have been done extensively in the studies of nuclear
symmetry energy at low densities
\cite{tsang09,shetty07,fami06,tsang04,chen05}.

To use $\pi^-/\pi^+$ to probe the high-density behavior of nuclear
symmetry energy has evident advantage within both the $\Delta$
resonance model and the statistical model \cite{Sto86,Ber80}.
Several hadronic transport models have quantitatively shown that
$\pi^-/\pi^+$ ratio is indeed sensitive to the symmetry energy
\cite{LiBA02,yong06,Gai04,LiQF05b}, especially around pion
production threshold. These transport models, however, usually use
different momentum dependent interactions among nucleons. The
importance of the momentum dependence of nuclear symmetry
potential was seldom mentioned in other transport models except
our transport model used in the present studies. In the framework
of the isospin-dependent Boltzmann-Uehling- Uhlenbeck transport
model, as an example, we studied the effect of the momentum
dependence of nuclear symmetry potential on $\pi^{-}/\pi ^{+}$ in
the neutron-rich reaction of $^{132}$Sn+$^{124}$Sn at a beam
energy of $400$ MeV/nucleon. It is found that the momentum
dependence of nuclear symmetry potential affects the compressed
density of colliding nuclei, numbers of produced $\pi^{-}$ and
$\pi ^{+}$, as well as the value of $\pi^{-}/\pi ^{+}$ ratio.


The isospin and momentum-dependent mean-field potential used in
the present work is \cite{Das03}
\begin{eqnarray}
U(\rho, \delta, \textbf{p},\tau)
=A_u(x)\frac{\rho_{\tau^\prime}}{\rho_0}+A_l(x)\frac{\rho_{\tau}}{\rho_0}\nonumber\\
+B\left(\frac{\rho}{\rho_0}\right)^\sigma\left(1-x\delta^2\right)\nonumber
-8x\tau\frac{B}{\sigma+1}\frac{\rho^{\sigma-1}}{\rho_0^\sigma}\delta\rho_{\tau^{\prime}}\nonumber\\
+\sum_{t=\tau,\tau^{\prime}}\frac{2C_{\tau,t}}{\rho_0}\int{d^3\textbf{p}^{\prime}\frac{f_{t}(\textbf{r},
\textbf{p}^{\prime})}{1+\left(\textbf{p}-
\textbf{p}^{\prime}\right)^2/\Lambda^2}},
\label{Un}
\end{eqnarray}
where $\rho_n$ and $\rho_p$ denote neutron ($\tau=1/2$) and proton
($\tau=-1/2$) densities, respectively.
$\delta=(\rho_n-\rho_p)/(\rho_n+\rho_p)$ is the isospin asymmetry
of nuclear medium. All parameters in the preceding equation can be
found in refs.\cite{IBUU04}. The variable $x$ is introduced to
mimic different forms of the symmetry energy predicted by various
many-body theories without changing any property of symmetric
nuclear matter and the value of symmetry energy at normal density
$\rho_0$. Because the purpose of present studies is just to see
how large the effect of momentum dependence of nuclear symmetry
potential on charged pion ratio, we let the variable $x$ be $1$.
In fact, behavior of nuclear symmetry energy at supra-densities is
still in controversy. For example, some authors concluded from the
FOPI data that the symmetry energy at supra-densities is very soft
\cite{xiao09}. Others concluded, however, the opposite situation
\cite{feng}, which is supported by the work of the Catania group.
With above choices the symmetry energy obtained from the preceding
single-particle potential is consistent with the Hartree-Fock
prediction using the original Gogny force \cite{Das03} and is also
favored by recent studies based on FOPI experimental data
\cite{xiao09}. The main characteristic of the present single
particle is the momentum dependence of nuclear symmetry potential,
which has evident effect on energetic free $n/p$ ratio in
heavy-ion collisions \cite{IBUU04}. But the momentum dependence of
nuclear symmetry potential on $\pi^{-}/\pi ^{+}$ ratio was seldom
reported. In this note we study the momentum dependence of nuclear
symmetry potential on $\pi^{-}/\pi ^{+}$ ratio. We keep the
isoscalar potential fixed while changing the symmetry potential
from momentum dependent symmetry potential to momentum independent
symmetry potential and keep the symmetry energy fixed
\cite{IBUU04}. The reaction channels related to pion production
and absorption are
\begin{eqnarray}
&& NN \longrightarrow NN, \nonumber\\ && NR \longrightarrow NR,
\nonumber\\ && NN \longleftrightarrow NR, \nonumber\\ && R
\longleftrightarrow N\pi,
\end{eqnarray}
where $R$ denotes $\Delta $ or $N^{\ast }$ resonances. In the
present work, we use the isospin-dependent in-medium reduced $NN$
elastic scattering cross section from the scaling model according
to nucleon effective mass \cite{factor,neg,pan,gale}. For
in-medium $NN$ inelastic scattering cross section, we use the
forms in free space since it is quite controversial.


\begin{figure}[th]
\begin{center}
\includegraphics[width=0.5\textwidth]{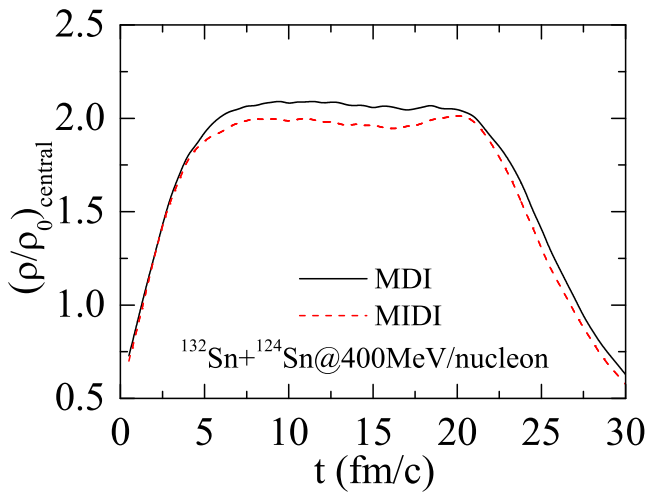}
\end{center}
\caption{(Color online) Evolution of the central baryon density
for the central reaction $^{132}$Sn+$^{124}$Sn at a beam energy of
400 MeV/nucleon with and without momentum dependence of nuclear
symmetry potential, signed with MDI and MIDI, respectively.}
\label{hdnz}
\end{figure}
Fig.~\ref{hdnz} shows the effect of the momentum dependence of
nuclear symmetry potential on the central baryon density of
colliding nuclei. It is seen that the maximum baryon density is
about 2 times normal nuclear matter density. Moreover, the
compression is sensitive to the momentum dependence of nuclear
symmetry potential. The momentum independence of nuclear symmetry
potential makes the nuclear matter less compressed whereas the
momentum dependence of nuclear symmetry potential causes a larger
compression.

\begin{figure}[th]
\begin{center}
\includegraphics[width=0.5\textwidth]{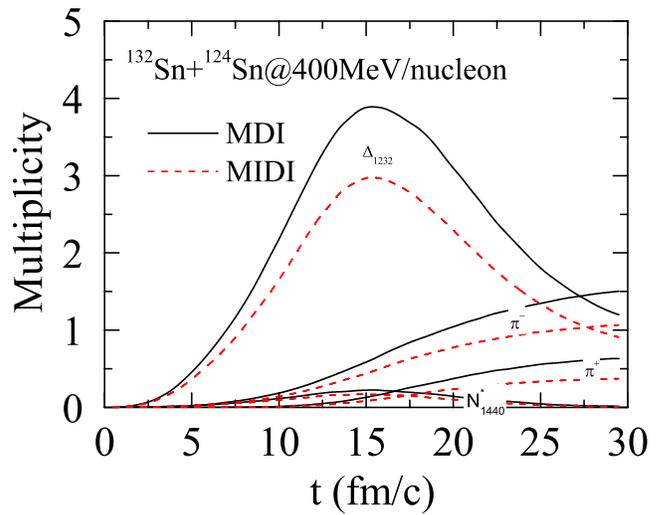}
\end{center}
\caption{(Color online) Evolution of $\pi^{-}$, $\pi ^{+}$ and
$\Delta(1232)$, $N^{*}(1440)$ multiplicities in the central
reaction $^{132}$Sn+$^{124}$Sn at a beam energy of 400 MeV/nucleon
with and without momentum dependence of nuclear symmetry
potential, respectively.} \label{multi}
\end{figure}
To see more clearly effect of the momentum dependence of nuclear
symmetry potential on pion production, we show in Fig.~\ref{multi}
the multiplicities of $\pi^+$, $\pi^-$ and $\Delta(1232)$, $N^{*}$
as a function of time. We can first see that, owing to small
compression with momentum independent symmetry potential shown in
Fig.~\ref{hdnz}, the momentum independence of nuclear symmetry
potential decreases the productions of resonances, especially
$\Delta(1232)$. More $\Delta(1232)$ resonances are produced than
$N^{*}$. This is because $N^{*}(1440)$ is related to more
energetic collisions. Second, we see that numbers of produced
charged pions are also reduced with the momentum independent
symmetry potential, especially for $\pi^-$. In the studies, the
usage of momentum dependence of nuclear symmetry potential
increases charged pions about 30\%. We also made simulations of
changing nuclear incompressibility ($\delta K\sim 20$) and find
that the isoscalar part of the Equation of State has little
effects on charged pion yields. But the momentum dependence of the
isoscalar part of the Equation of State also has evident affection
on pion yields \cite{liq06}. All these results indicate the
importance of the momentum dependence of nuclear potential on the
studies of pion production.

\begin{figure}[th]
\begin{center}
\includegraphics[width=0.5\textwidth]{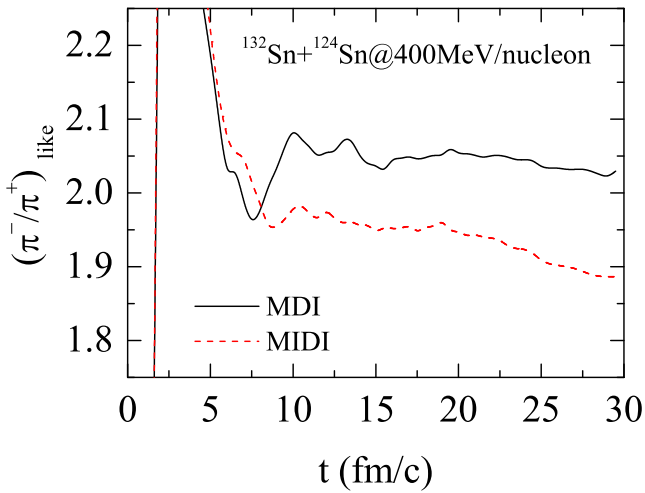}
\end{center}
\caption{(Color online) Evolution of the $\pi^-/\pi^+$ ratio in
the reaction $^{132}$Sn+$^{124}$Sn at a beam energy of 400
MeV/nucleon with and without momentum dependence of nuclear
symmetry potential.} \label{tratio}
\end{figure}
Shown in Fig.~\ref{tratio} is effect of the momentum dependence of
nuclear symmetry potential on the $(\pi^-/\pi^+)_{like}$ ratio as
a function of time in the central reaction $^{132}$Sn+$^{124}$Sn
at a beam energy of $400$ MeV/nucleon. In the dynamics of pion
resonance productions and decays the $(\pi^-/\pi^+)_{like}$ ratio
reads
\begin{equation}
(\pi^-/\pi^+)_{like}\equiv
\frac{\pi^-+\Delta^-+\frac{1}{3}\Delta^0+\frac{2}{3}N^{*0}}
{\pi^++\Delta^{++}+\frac{1}{3}\Delta^++\frac{2}{3}N^{*+}}.
\end{equation}
This ratio naturally becomes $\pi^-/\pi^+$ ratio at the freeze-out
stage. From Fig.~\ref{tratio} we can first see that sensitivity of
$(\pi^-/\pi^+)_{like}$ ratio to the effect of the momentum
dependence of nuclear symmetry energy is clearly shown after $t=10
fm/c$. With the momentum dependent symmetry potential (MDI),
$\pi^-/\pi^+$ ratio is higher than that with the momentum
independent symmetry potential (MIDI), the effect of the momentum
dependence of nuclear symmetry potential in this study is about
7.4\%. The $\pi^-/\pi^+$ ratio with the momentum dependent
symmetry potential is higher than that with the momentum
independent symmetry potential is consistent with the free n/p
ratio's momentum dependence of nuclear symmetry potential
\cite{IBUU04}. With the momentum dependent symmetry potential,
free n/p ratio is lower than that with the momentum independent
symmetry potential. The n/p ratio of dense matter thus has a
opposite situation, i.e., with the momentum dependent symmetry
potential, \emph{dense} matter's n/p ratio is higher than that
with the momentum independent symmetry potential. According to
both the $\Delta$ resonance model and the statistical model
\cite{Sto86,Ber80}, with the momentum dependent symmetry
potential, $\pi^-/\pi^+$ ratio is higher than that with the
momentum independent symmetry potential.

Therefore, if the published PRL paper \cite{xiao09} of Xiao et al.
uses a momentum independent symmetry potential, according to the
experimental data, the resulting symmetry energy will be more
soft. The whole physical result of that PRL paper \cite{xiao09}
still does not change, just more soft. To study the momentum
dependence of nuclear symmetry potential, high p$_{t}$ or kinetic
energy's neutron to proton ratio or light over heavy cluster's n/p
ratio may be useful \cite{IBUU04,yong105}.


In conclusion, based on the isospin-dependent
Boltzmann-Uehling-Uhlenbeck transport model, effect of the
momentum dependence of nuclear symmetry potential on $\pi^{-}/\pi
^{+}$ ratio in the neutron-rich reaction $^{132}$Sn+$^{124}$Sn at
a beam energy of $400$ MeV/nucleon is studied. It is found that
momentum dependent nuclear symmetry potential increases the
compressed density of colliding nuclei, numbers of produced
resonances $\Delta(1232)$, $N^{*}(1440)$, $\pi^{-}$ and $\pi
^{+}$, as well as the value of $\pi^{-}/\pi ^{+}$ ratio. It is
therefore necessary to consider the momentum dependence of nuclear
symmetry potential while studying the effect of nuclear symmetry
energy by using heavy-ion collisions.


The author Y. Gao thanks Prof. Bao-An Li for providing the code
and useful guidance while he stayed at Institute of Modern
Physics, Chinese Academy of Sciences. This work is supported in
part by the National Natural Science Foundation of China under
grants 10740420550, 10875151, 10947109, 10975064 and 11005157.


\begin{thebibliography}{00}

\bibitem{xiao09}Z.G. Xiao, B.A. Li, L.W. Chen, G.C. Yong, M. Zhang, Phys. Rev.
Lett. {\bf 102} 062502 (2009).

\bibitem{ditoro1}M. Di Toro, V. Baran, M. Colonna, V. Greco, J. Phys. G: Nucl.
Phys. {\bf 37}, 083101 (2010).

\bibitem{xu09}J. Xu, C.M. Ko, Y. Oh, Phys. Rev. {\bf C81}, 024910 (2010).

\bibitem{Rei07}W. Reisdorf, M. Stockmeier, A. Andronic, M.L. Benabderrahmane,
O.N. Hartmann, N. Herrmann, K.D. Hildenbrand, Y.J. Kima, et al.,
Nucl. Phys. {\bf A781}, 459 (2007).

\bibitem{yong06}G.C. Yong, B.A. Li, L.W. Chen, W. Zuo, Phys. Rev. {\bf C73}, 034603 (2006).

\bibitem{yong101}G.C. Yong, Phys. Rev. {\bf C82}, 064604 (2010).

\bibitem{yong102}G.C. Yong, arXiv: 1012.2921 (2010).

\bibitem{LiBA02}B.A. Li, Phys. Rev. Lett. {\bf 88}, 192701 (2002).

\bibitem{Bro00}B.A. Brown, Phys. Rev. Lett. {\bf 85}, 5296 (2000).

\bibitem{Dan02a}P. Danielewicz, R. Lacey, W.G. Lynch, Science {\bf 298}, 1592 (2002).

\bibitem{Bar05}V. Baran, M. Colonna, V. Greco, M. Di Toro, Phys. Rep. {\bf 410}, 335 (2005).

\bibitem{LCK08}B.A. Li, L.W. Chen and C.M. Ko, Phys. Rep. {\bf 464}, 113 (2008).

\bibitem{Sum94}K. Sumiyoshi and H. Toki, Astrophys. J. {\bf 422}, 700 (1994).

\bibitem{Lat04}J.M. Lattimer, M. Prakash, Science {\bf 304}, 536 (2004).

\bibitem{Ste05a}A.W. Steiner, M. Prakash, J.M. Lattimer, P.J. Ellis, Phys. Rep. {\bf 411}, 325 (2005).

\bibitem{Kut94}M. Kutschera, Phys. Lett. {\bf B340}, 1 (1994).

\bibitem{Kub99}S. Kubis and M. Kutschera, Acta Phys. Pol. {\bf B30},
2747 (1999); Nucl. Phys. {\bf A720}, 189 (2003).

\bibitem{Che07}L.W. Chen, C.M. Ko, B.A. Li, Phys. Rev. {\bf C76}, 054316 (2007).

\bibitem{LiZH06}Z.H. Li, U. Lombardo, H.J. Schulze, W. Zuo, L.W. Chen, and H.R. Ma, Phys. Rev. {\bf C74}, 047304 (2006).

\bibitem{Pan72}V.R. Pandharipande, V.K. Garde, Phys. Lett. {\bf B39}, 608 (1972).

\bibitem{Fri81}B. Friedman, V.R. Pandharipande, Nucl. Phys. {\bf A361}, 502 (1981).

\bibitem{Wir88a}R.B. Wiringa, V. Fiks, and A. Fabrocini, Phys. Rev. {\bf C38}, 1010 (1988).

\bibitem{Kra06}P. Krastev and F. Sammarruca, and Sammarruca. F, Phys. Rev. {\bf C74}, 025808 (2006).

\bibitem{Szm06}A. Szmaglinski, W. W\'{o}jcik, M. Kutschera, Acta Phys. Polon. {\bf B37}, 227 (2006).

\bibitem{Cha97}E. Chabanat, P. Bonche, P. Haensel, J. Meyer, R. Schaeffer, Nucl. Phys. {\bf A627} (1997) 710; {\it ibid}, 635 (1998) 231.

\bibitem{Sto03}J.R. Stone, J.C. Miller, R. Koncewicz, P. D. Stevenson, M. R. Strayer, Phys. Rev. {\bf C68}, 034324 (2003).

\bibitem{Che05b}L.W. Chen, C.M. Ko and B.A. Li, Phys. Rev. {\bf C72}, 064309 (2005).

\bibitem{Dec80}J. Decharge and D. Gogny, Phys. Rev. {\bf C21}, 1568 (1980).

\bibitem{MS}W.D. Myers and W.J. Swiatecki, Acta Phys. Pol. {\bf B26},111 (1995).

\bibitem{Kho96}D.T. Khoa, W. von Oertzen, A.A. Ogloblin, Nucl. Phys. {\bf A602}, 98 (1996).

\bibitem{Bas07}D.N. Basu, T. Mukhopadhyay, Acta Phys. Polon. {\bf B38}, 169 (2007).

\bibitem{Ban00}S. Banik and D. Bandyopadhyay, J. Phys. G {\bf 26}, 1495 (2000).

\bibitem{yong105}G.C. Yong, Phys. Rev. {\bf C81}, 054603 (2010).

\bibitem{tsang09}M.B. Tsang, Y.X Zhang, P. Danielewicz, M. Famiano, Z.X. Li, W.G. Lynch, and A.W. Steiner, Phys. Rev. Lett. {\bf 102}, 122701 (2009).

\bibitem{shetty07}D.V. Shetty, S. J. Yennello, and G. A. Souliotis, Phys. Rev. {\bf C76}, 024606 (2007).

\bibitem{fami06}M.A. Famiano, T. Liu, W.G. Lynch, M. Mocko, A.M. Rogers, M.B. Tsang, M.S. Wallace, R.J. Charity, S. Komarov, D.G. Sarantites, L.G. Sobotka, and G. Verde, Phys. Rev. Lett. {\bf 97}, 052701 (2006).

\bibitem{tsang04}M.B. Tsang, T.X. Liu, L. Shi, P. Danielewicz, C.K. Gelbke, X.D. Liu, W.G. Lynch, W.P. Tan, G. Verde, A. Wagner, and H.S. Xu, Phys. Rev. Lett. {\bf 92}, 062701 (2004).

\bibitem{chen05}L.W. Chen, C.M. Ko and B.A. Li, Phys. Rev. Lett. {\bf 94}, 032701 (2005).

\bibitem{Sto86}R. Stock, Phys. Rep., {\bf 135}, 259 (1986).

\bibitem{Ber80}G.F. Bertsch, Nature {\bf 283}, 280 (1980);
A. Bonasera and G.F. Bertsch, Phys. Let. B195 (1987) 521.

\bibitem{Gai04}T. Gaitanos, M. Di Toro, S. Typel, V. Baran, C. Fuchs, V. Greco, H.H. Wolter, Nucl. Phys. {\bf A732}, 24 (2004).

\bibitem{LiQF05b}Q.F. Li, Z.X. Li, S. Soff, M. Bleicher, and Horst St\"{o}cker, Phys. Rev. {\bf C72}, 034613 (2005).

\bibitem{Das03}C. B. Das, S. Das Gupta, C. Gale, and B.A. Li, Phys. Rev. {\bf C67}, 034611 (2003).

\bibitem{IBUU04}B.A. Li, C.B. Das, S. Das Gupta, C. Gale, Nucl. Phys. {\bf A735}, 563 (2004); Phys. Rev. {\bf C69}, 064602
(2004).

\bibitem{feng}Z.Q. Feng and G.M. Jin , Phys. Lett. {\bf B683}, 140 (2002).

\bibitem{factor}B.A. Li and L.W. Chen, Phys. Rev. {\bf C72}, 064611 (2005).

\bibitem{neg}J.W. Negele and K. Yazaki, Phys. Rev. Lett. {\bf 47}, 71
(1981).

\bibitem{pan}V.R. Pandharipande and S.C. Pieper, Phys. Rev. {\bf C45}, 791
(1992).

\bibitem{gale}D. Persram and C. Gale, Phys. Rev. {\bf C65}, 064611 (2002).

\bibitem{liq06}Q. Li, Z. Li, S. Soff, M. Bleicher and H. St\"{o}cker,
J. Phys. G: Nucl. Part. Phys. {\bf 32}, 151 (2006).

\end{thebibliography}
\end{document}